\documentclass{article}%
\usepackage{amsmath}
\usepackage{graphicx}
\usepackage{amsfonts}
\usepackage{amssymb}%
\begin{document}

\begin{center}
{\LARGE First Order Actions: a New View}

\bigskip

M. C. Bertin$^{(a)}$\footnote{mcbertin@ift.unesp.br}, B. M. Pimentel$^{(a)}%
$\footnote{pimentel@ift.unesp.br}, P. J. Pompeia$^{(a,b)}$%
\footnote{pompeia@ift.unesp.br}

\bigskip

$^{(a)}$ Instituto de F\'{\i}sica Te\'{o}rica - Universidade Estadual
Paulista,\newline Rua Pamplona 145, 01405-900, S\~{a}o Paulo, SP,
Brazil.\newline$^{(b)}$ Centro T\'{e}cnico Aeroespacial - Instituto de Fomento
e Coordena\c{c}\~{a}o Industrial - Divis\~{a}o de Confiabilidade
Metrol\'{o}gica Aeroespacial\newline Pra\c{c}a Mal. Eduardo Gomes, 50,
12228-901, S\~{a}o Jos\'{e} dos Campos, SP, Brazil.\newline

\bigskip


\bigskip


\bigskip

\textbf{Abstract:} \textit{We analyse systems described by first order actions
using the Hamilton-Jacobi (HJ) formalism for singular systems. In this study
we verify that generalized brackets appear in a natural way in HJ approach,
showing us the existence of a symplectic structure in the phase space of this
formalism.}
\end{center}

\section{Introduction}

Systems described by first order actions, \textit{i.e.} Lagrangians linear in
the velocities \cite{Schwinger, Symanzik} appear in many branches in Physics.
In 1928 Dirac proposed a system with first order action \cite{Dirac-eq} which
has been used since then to describe fermion fields. Later in the 1930's, a
Lagrangian with a linear kinematic term was also proposed in order to describe
bosonic fields (DKP\ theory) \cite{DKP, DKP A, DKP B}. In gravitation, this
type of Lagrangian appeared for the first time in 1919 with Palatini's work
\cite{Palatini}, which came to be the basis of a new method of variation.
First order actions were also present in Schwinger's development of quantum
theory \cite{Schwinger-FT, Schwinger-FT 2}.

A significant feature of these systems is that they always have a null Hessian
matrix, \textit{i.e.} they are singular (constrained) systems, and hence they
must be properly treated in order to accomplish a hamiltonian formulation.

In the particular case of first-order actions, different approaches can be
applied. The most usual one is the formalism developed in 1950 by Dirac
\cite{Dirac0, Dirac1, Dirac2} to treat general constrained systems, in which
the hamiltonian structure is employed \cite{Ham-Struc, Ham-Struc1, Ham-Struc2,
Ham-Struc3, Ham-Struc4}. One of the main features of the Dirac formalism is
the fact that it allows one to introduce generalized brackets which conduct to
a consistent quantization of the system. The application of this formalism to
first-order action can be found in reference \cite{Govaerts}.

Another approach to deal with Lagrangians with linear velocities was developed
by Faddeev and Jackiw \cite{F-J} in 1988, where generalized brackets framed on
the symplectic structure of phase space were also introduced, leading to a
consistent quantization at least to purely bosonic variables.

A third approach that can also be applied to study such systems is the
Hamilton-Jacobi (HJ) formalism, which is based on the Carath\'{e}odory
Equivalent Lagrangian method \cite{Carat}. This method, developed by
Carath\'{e}odory to treat regular systems with first derivatives, is an
alternative way to obtain the Hamilton-Jacobi equation starting from
lagrangian formalism. In 1992 G\"{u}ler generalized Carath\'{e}odory's method
to treat singular systems \cite{Guler 1, Guler 2} and more recently Pimentel
and Teixeira \cite{Rand Pim1, Rand Pim2}, worked with Lagrangians with higher
order derivatives. In 1998 Pimentel, Teixeira and Tomazelli made an extension
to deal with Berezinian singular systems \cite{Rand Pim Jef}. Important
applications of this method can be found in literature \cite{REF HJ, REF HJ 1,
REF HJ 2, REF HJ 3,REF HJ 4}, including an application to Lagrangians linear
in the velocities \cite{Guler -lin-veloc}, where no generalized brackets are introduced.

In this work we intend to study systems with first-order actions \textit{via}
Hamilton-Jacobi formalism and show how generalized brackets and a symplectic
structure appear in a natural way. Accordingly, in the two next sections we
will make a review of first-order actions and HJ formalism, respectively.
Afterwards we will apply the HJ structure to the Lagrangians of interest and
see how generalized brackets are introduced. Then we will show some examples
and\ at last some concluding remarks will be made.

\section{First Order Actions\label{Sec1}}

We shall consider here a system whose dynamical evolution is described by some
variational principle from the action integral%
\begin{equation}
\mathcal{A}\left[  z_{A}\right]  =\int_{i_{i}}^{t_{f}}dtL\left(  z_{A},\dot
{z}_{A}\right)  ,\,\,\,\,\,\,\,\,\,\,\,\,\,\,\,\,\,\,\,A=1,...,N. \label{acao}%
\end{equation}
In this expression $z_{A}$ are coordinates and $\dot{z}_{A}$ are the time
derivative of $z_{A}\,$. We will assume that the Lagrangian function has
linear dependence on the velocities $\dot{z}_{A}$, \textit{i.e.},%
\begin{equation}
L\left(  z_{A},\dot{z}_{A}\right)  =\dot{z}_{A}K^{A}\left(  z_{B}\right)
-V\left(  z_{B}\right)  , \label{L}%
\end{equation}
where $K^{A}$ and $V$ are arbitrary functions of the coordinates $z_{B}$.

One can immediately verify that the variational problem remains the same if we
consider, instead of $L$, another Lagrangian function $\tilde{L}\left(
z_{A},\dot{z}_{A},t\right)  $, which differs from the first by a total time
derivative:%
\begin{align}
\tilde{L}\left(  z_{A},\dot{z}_{A},t\right)   &  =L\left(  z_{A},\dot{z}%
_{A}\right)  +\frac{d}{dt}Y\left(  z_{A},t\right)  =\nonumber\\
&  =L+\frac{\partial Y}{\partial t}+\dot{z}_{A}\frac{\partial Y}{\partial
z_{A}}. \label{L_transf}%
\end{align}
An interesting property of this transformation in $L$ is that we can preserve
the structure of (\ref{L}) for $\tilde{L}$, $\tilde{L}\left(  z,\dot
{z},t\right)  =\dot{z}_{A}\tilde{K}^{A}\left(  z,t\right)  -\tilde{V}\left(
z,t\right)  ,$ if $K^{A}$ and $V$ transform respectively as%
\begin{align}
\tilde{K}^{A}\left(  z\right)   &  =K^{A}\left(  z\right)  +\frac{\partial
Y}{\partial z_{A}}\left(  z,t\right)  ,\nonumber\\
\tilde{V}\left(  z\right)   &  =V\left(  z\right)  -\frac{\partial Y}{\partial
t}. \label{K_transf}%
\end{align}

Since the variational problem is unchanged by the transformations above, we
can expect that the equations of motion depend on quantities that are
invariant by these transformations. One example of a quantity that has such a
property is the {}``curl'' of $K^{A}$:%
\begin{equation}
M^{AB}\equiv\frac{\partial K^{B}}{\partial z_{A}}-\frac{\partial K^{A}%
}{\partial z_{B}}=-M^{BA}=\tilde{M}^{AB}, \label{M}%
\end{equation}
and as it will be seen in the next section, it is related to equations of motions.

\section{The Hamilton-Jacobi Formalism for an Arbitrary Lagrangian}

The Hamilton-Jacobi (HJ) equation is usually obtained from the hamiltonian
approach when a specific canonical transformation is considered. An
alternative path to reach HJ formalism was developed by
Carath\'{e}odory\cite{Carat} and starts from the lagrangian approach without
mentioning the hamiltonian one. According to Carath\'{e}odory Equivalent
Lagrangian method, a minimum of the action $\mathcal{A}$ can be found when we
consider a set of functions $\beta_{A}\left(  z_{B},t\right)  $ such that
\begin{equation}
\tilde{L}\left(  z_{A},\dot{z}_{A}=\beta_{A}\left(  z_{B},t\right)  \right)
=L\left(  z_{A},\dot{z}_{A}\right)  +\frac{\partial}{\partial t}Y\left(
z_{A},t\right)  +\frac{\partial Y\left(  z_{A},t\right)  }{\partial z_{A}}%
\dot{z}_{A}=0, \label{L'=0}%
\end{equation}
and in a neighbourhood of $\dot{z}_{A}=\beta_{A}\left(  z_{B},t\right)  $ the
condition $\tilde{L}\left(  z_{A},\dot{z}_{A}\right)  >0$ is satisfied. From
these conditions it follows that%
\begin{equation}
p^{A}\equiv\left.  \frac{\partial L}{\partial\dot{z}_{A}}\right|  _{\dot
{z}_{B}=\beta_{B}}=-\left.  \frac{\partial Y}{\partial z_{A}}\right|
_{\dot{z}_{B}=\beta_{B}}. \label{momenta}%
\end{equation}

If $L$ is a singular Lagrangian then the Hessian matrix, $H^{AB}%
=\frac{\partial^{2}L}{\partial\dot{z}_{A}\partial\dot{z}_{B}}$, has a null
determinant, $\det H^{AB}=0$; and if this matrix has rank $P=N-R$, then we can
find a $P\mathsf{x}P$ submatrix such that%
\begin{equation}
\det H^{ab}=\det\frac{\partial p^{a}}{\partial\dot{z}_{b}}\neq
0,~\,\,\,\,\,\,\,\,\,\,\,\,a,b=R+1,...,N. \label{subHess}%
\end{equation}

For this case we can verify that $R$ momenta $p^{\alpha}$ ($\alpha=1,...R$)
have no dependence on any velocity, which means that $R$ velocities cannot be
written as functions of $z$ and $p$, as it happens to $\dot{z}_{b}$, $\dot
{z}_{b}=f_{b}\left(  z_{A},p^{a}\right)  $. We conclude that $R$ relations of
the type%
\begin{equation}
p^{\alpha}=-H^{\alpha}\left(  t,z_{\beta}\equiv t_{\beta},z_{a},p^{a}\right)
\label{vinc1}%
\end{equation}
must be satisfied.

Moreover if we define $H_{0}\equiv p^{A}\dot{z}_{A}-L$ it follows from
(\ref{L'=0}) that%
\begin{equation}
p^{0}+H^{0}\left(  t,t_{\alpha},z_{a},p^{a}\right)  =0, \label{eqHJ}%
\end{equation}
where $p^{0}\equiv\frac{\partial S}{\partial t}$.

We see that equations (\ref{vinc1}) and (\ref{eqHJ}) lead us to define $R+1$
conditions%
\begin{equation}
\phi^{\alpha^{\prime}}\equiv p^{\alpha^{\prime}}+H^{\alpha^{\prime}}\left(
t_{\beta^{\prime}},z_{a},p^{a}\right)  =0,~\,\,\,\,\,\,\,\,\alpha^{\prime
},\beta^{\prime}=0,1,...,R, \label{vinculos}%
\end{equation}
where $t^{0}\equiv t$. These conditions $\phi^{\alpha^{\prime}}=0$ are usually
called \textit{constraints}, and they constitute a set of first order partial
differential equations, called \textit{Hamilton-Jacobi Partial Differential
Equations} (HJPDE).

\subsection{Integrability Conditions}

In order to integrate the HJPDE (\ref{vinculos}) we can use the method of
characteristics\cite{Carat}, which conducts us to total differential equation%
\begin{equation}
\left\{
\begin{array}
[c]{c}%
d\eta^{I}=E^{IJ}\frac{\partial\phi^{\alpha}}{\partial\eta^{J}}dt_{\alpha
}=\left\{  \eta^{I},\phi^{\alpha}\right\}  dt_{\alpha},\\
dS=\frac{\partial S}{\partial z_{A^{\prime}}}dz_{A^{\prime}}=p^{A^{\prime}%
}\frac{\partial\phi^{\alpha}}{\partial p^{A^{\prime}}}dt_{\alpha},
\end{array}
\right.  \,\,\
\begin{array}
[c]{c}%
I,J=\left(  \zeta;A^{\prime}\right)  ,\,\zeta=1,2;\\
\alpha=0,...,R;~A^{\prime}=0,...,N;
\end{array}
\label{eq carac}%
\end{equation}
where $\left\{  \eta^{1A\prime}\right\}  =\left\{  z_{A^{\prime}}\right\}  $
and $\left\{  \eta^{2A^{\prime}}\right\}  =\left\{  p^{A^{\prime}}\right\}  $,
$E^{IJ}=\delta_{\,B^{\prime}}^{A^{\prime}}\left[  \delta_{\,1}^{\zeta}%
\delta_{\,\sigma}^{2}-\delta_{\,2}^{\zeta}\delta_{\,\sigma}^{1}\right]  ,$
$(I=\left(  \zeta,A^{\prime}\right)  ,\,J=\left(  \sigma,\,B^{\prime}\right)
)$. In this expression we use the definition of Poisson Brackets $\left\{
F,G\right\}  =\frac{\partial F}{\partial\eta^{I}}E^{IJ}\frac{\partial
G}{\partial\eta^{J}}$. According to this method, if the characteristic
equations are integrable then the HJPDE will have a unique solution
(determined by initial conditions). To obtain (\ref{eq carac}) we assume that
the momenta and coordinates are independent quantities, and we can observe
that if $d\eta^{I}$ (whose equations will be called equations of motions)
constitute an integrable system, then $dS$ will be integrable as a consequence.

To assure the integrability of the equations of motion we must recall from the
theory of differential equations that, associated with a set of total
equations, $dx_{I}=b_{I}^{\alpha}\left(  x_{J}\right)  dt_{\alpha}$, there are
linear operators $X^{\alpha}$ such that%
\begin{equation}
X^{\alpha}F\left(  x^{J}\right)  =b_{I}^{\alpha}\frac{\partial F}{\partial
x_{I}}=0. \label{eq dif parc}%
\end{equation}
From this result it is obvious that%
\begin{equation}
\left[  X^{\alpha},X^{\beta}\right]  F=\left(  X^{\alpha}X^{\beta}-X^{\beta
}X^{\alpha}\right)  F=0, \label{comut F}%
\end{equation}
if $F$ is at least twice differentiable.

The partial differential equations $X_{\alpha}F=0$ are said to be
\textit{complete} if
\[
\left[  X^{\alpha},X^{\beta}\right]  F=C_{~\,\,\gamma}^{\alpha\beta}X^{\gamma
}F.
\]
If this condition is not satisfied we can define a new operator $X$ such that
$XF=0$, and it must be added to previous set $X_{\alpha}$, and we must verify
if this new set is complete. This procedure must be repeated until a complete
set is obtained. The total differential equations will be integrable when the
associated partial equations constitute a complete set.

We must notice that when we define a new operator $X$ we are imposing a
restriction to phase space. In fact we are searching a subspace of the
original phase space where the equations of motions can be integrated.

Considering now the specific case of the equations of motion (\ref{eq carac})
we have $X^{\alpha}F=\left\{  F,\phi^{\alpha}\right\}  $, and using Jacobi
identity for Poisson Brackets it follows $\left[  X^{\alpha},X^{\beta}\right]
F=-\left\{  F,\left\{  \phi^{\alpha},\phi^{\beta}\right\}  \right\}  $. The
equations of motion will be integrable if%
\begin{equation}
\left\{  \phi^{\alpha},\phi^{\beta}\right\}  =C_{\,\,\gamma}^{\alpha\beta}%
\phi^{\gamma}=0, \label{cond int0}%
\end{equation}
or considering the independence of $t_{\alpha}$,%
\begin{equation}
d\phi^{\alpha}=\left\{  \phi^{\alpha},\phi^{\beta}\right\}  dt_{\beta}=0.
\label{cond int1}%
\end{equation}

If these conditions are not satisfied we must restrict our phase space with
new relations $\phi=0$ until a complete set of partial differential equations
is obtained.

\section{The HJ Formalism for First Order Actions}

Let us now consider the specific case of section (\ref{Sec1}). According to
the previous section when the condition $\tilde{L}=0$ and the action is a
minimum, the momenta canonically conjugated to $z_{A}$ and $z_{0}=t$ are
respectively%
\begin{align}
p^{A}  &  =-\frac{\partial Y}{\partial z_{A}}%
,\,\,\,\,\,\,\,\,\,\,\,\,\,\,\,\,\,\,\,\,\,A=1,...N,\nonumber\\
p^{0}  &  =-\frac{\partial Y}{\partial t}. \label{momentos}%
\end{align}
However when $\tilde{L}=0$, we have%
\begin{equation}
\tilde{L}=\dot{z}_{A}\left(  K^{A}\left(  z\right)  -p^{A}\right)  -\dot
{z}_{0}\left(  V\left(  z\right)  +p^{0}\right)  =\dot{z}_{A^{\prime}}%
\tilde{K}^{A^{\prime}}\left(  z,t\right)  =0,\,\,\,\,A^{\prime}=0,1,...,N,
\label{L_transf=0}%
\end{equation}
where%
\begin{equation}
\tilde{K}^{0}\equiv-\left(  V\left(  z\right)  -\frac{\partial Y}{\partial
t}\left(  z,t\right)  \right)  . \label{K0}%
\end{equation}
If we consider $\dot{z}_{A^{\prime}}$ as independent quantities then%
\begin{equation}
\tilde{K}^{A^{\prime}}\left(  z,t\right)  =0\Rightarrow\left\{
\begin{array}
[c]{c}%
K^{A}\left(  z\right)  -p^{A}=0,\\
V\left(  z\right)  +p^{0}=0.
\end{array}
\right.  \label{EDPHJ_1ord}%
\end{equation}

This is the set of HJPDE, which leads us to define the constraints
\begin{equation}
\phi^{A^{\prime}}\equiv p^{A^{\prime}}-K^{A^{\prime}}\left(  z\right)  =0,
\label{Vinc_1ord}%
\end{equation}
with $K^{0}\equiv-V\left(  z\right)  $.\ From this result we see that all the
coordinates have the status of parameters and then, to be consistent with the
notation of the previous section, we define%
\begin{align*}
t_{A}  &  \equiv z_{A},\\
t_{0}  &  \equiv z_{0}\equiv t.
\end{align*}

\subsection{Integrability Conditions}

To test the integrability conditions we must calculate the total differential
of the constraints (\ref{Vinc_1ord}):%
\[
d\phi^{A^{\prime}}=\left\{  \phi^{A^{\prime}},\phi^{B^{\prime}}\right\}
dt_{B^{\prime}}=\left\{  \phi^{A^{\prime}},\phi^{0}\right\}  dt_{0}+\left\{
\phi^{A^{\prime}},\phi^{B}\right\}  dt_{B},
\]
or explicitly
\begin{equation}
d\phi^{0}=\left\{  \phi^{0},\phi^{B}\right\}  dt_{B}, \label{d_fi_zero}%
\end{equation}%
\begin{align}
d\phi^{A}  &  =\left\{  \phi^{A},\phi^{0}\right\}  dt_{0}+\left\{  \phi
^{A},\phi^{B}\right\}  dt_{B}=\nonumber\\
&  =\left\{  \phi^{A},\phi^{0}\right\}  dt_{0}+M^{AB}dt_{B}. \label{de_fi_A}%
\end{align}
If we consider the independence of the parameters $dt_{B^{\prime}}$ then we
see that the equations of motion are integrable only if $\left\{  \phi
^{A},\phi^{0}\right\}  =0,~M^{AB}=0$. If this is not the case we have some
problems, because all the coordinates are already parameters and no further
restriction can be done, \textit{i.e.} we cannot define a new constraint
$\phi=0$. And the obvious conclusion is: the system is not integrable.

Of course this analysis is valid when we consider $dt_{B^{\prime}}$ as
independent quantities. The question naturally arises: can we find a subspace
of the parameters space where the system becomes integrable? To answer this
question we admit that $M^{AB}$ is not nule and that such construction can be
done. Hence, in this subspace we have%
\begin{align*}
d\phi^{0}  &  =0,\\
d\phi^{A}  &  =0.
\end{align*}
The last expression shows
\begin{equation}
d\phi^{A}=\left\{  \phi^{A},\phi^{0}\right\}  dt_{0}+M^{AB}dt_{B}=0\Rightarrow
M^{AB}dt_{B}=-\left\{  \phi^{A},\phi^{0}\right\}  dt_{0}, \label{Mdt_B=PP_dt}%
\end{equation}
and the dependence among $dt_{B}$ and $dt_{0}$ becomes clear.

\subsubsection{The $M^{AB}$ Regular Case}

Let us now consider the case when $M^{AB}$ is a regular matrix. Hence
$\det\left(  M^{AB}\right)  \neq0$ and the matrix $M_{AB}^{-1}$ does exist. In
this case it is straightforward to verify that%
\begin{equation}
dt_{B}=-M_{BA}^{-1}\left\{  \phi^{A},\phi^{0}\right\}  dt_{0}. \label{dt_B}%
\end{equation}

If we substitute this result in (\ref{d_fi_zero}) it follows%
\[
d\phi^{0}=-\left\{  \phi^{0},\phi^{B}\right\}  M_{BA}^{-1}\left\{  \phi
^{A},\phi^{0}\right\}  dt_{0}=\left\{  \phi^{B},\phi^{0}\right\}  M_{BA}%
^{-1}\left\{  \phi^{A},\phi^{0}\right\}  dt_{0},
\]
and since $M_{BA}^{-1}=-M_{AB}^{-1}$ it is immediate that $d\phi^{0}=0$,
because $\left\{  \phi^{B},\phi^{0}\right\}  \left\{  \phi^{A},\phi
^{0}\right\}  =\left\{  \phi^{A},\phi^{0}\right\}  \left\{  \phi^{B},\phi
^{0}\right\}  $.

Now considering the dependence stablished by (\ref{dt_B}), the differential of
any function $E=E\left(  z,p\right)  $ is given by%
\[
dE=\left[  \left\{  E,\phi^{0}\right\}  -\left\{  E,\phi^{B}\right\}
M_{BA}^{-1}\left\{  \phi^{A},\phi^{0}\right\}  \right]  dt_{0}.
\]
We can now introduce new Brackets%
\begin{equation}
\left\{  F,G\right\}  _{\ast}\equiv\left\{  F,G\right\}  -\left\{  F,\phi
^{B}\right\}  M_{BA}^{-1}\left\{  \phi^{A},G\right\}  , \label{PD_HJ}%
\end{equation}
such that%
\begin{equation}
dE=\left\{  E,\phi^{0}\right\}  _{\ast}dt_{0}. \label{dE}%
\end{equation}

In particular if we consider in (\ref{PD_HJ}) functions $F=F\left(
z_{A}\right)  $ and $G=G\left(  z_{B}\right)  $, then%
\begin{equation}
\left\{  F,G\right\}  _{\ast}=\frac{\partial F}{\partial z_{A}}M_{AB}%
^{-1}\frac{\partial G}{\partial z_{B}}, \label{Est_Simp}%
\end{equation}
and if $F=z_{A}$ and $G=z_{B}$,%
\begin{equation}
\left\{  z_{A},z_{B}\right\}  _{\ast}=M_{AB}^{-1}. \label{Est_Simp1}%
\end{equation}

Equations (\ref{Est_Simp}) and (\ref{Est_Simp1}) show there is a symplectic
structure in phase space in HJ approach.

We can still verify the consistence of this construction by taking $E=z_{C}$
in (\ref{dE}) and see if (\ref{dt_B}) is obtained:%
\begin{align*}
dz_{C}  &  =\left\{  z_{C},\phi^{0}\right\}  _{\ast}dt_{0}=\left[  \left\{
z_{C},\phi^{0}\right\}  -\left\{  z_{C},\phi^{B}\right\}  M_{BA}^{-1}\left\{
\phi^{A},\phi^{0}\right\}  \right]  dt_{0}=\\
&  =-\delta_{C}^{B}M_{BA}^{-1}\left\{  \phi^{A},\phi^{0}\right\}
dt_{0}=-M_{CA}^{-1}\left\{  \phi^{A},\phi^{0}\right\}  dt_{0}.
\end{align*}
If we consider that $z_{C}=t_{C}$ then the verification is straightforward.

At last, expliciting $\left\{  \phi^{A},\phi^{0}\right\}  $, we see%
\begin{equation}
dz_{C}=M_{CA}^{-1}\frac{\partial V}{\partial z_{A}}dt_{0}. \label{eq_mov_reg}%
\end{equation}
This result is in agreement to that one presented in \cite{Govaerts}, where
the lagrangian and hamiltonian approach are considered.

\subsubsection{The $M^{AB}$ Singular Case}

In what follows we will consider the case when $M^{AB}$ is a singular matrix
(\textit{i.e.} $\det\left(  M^{AB}\right)  =0$) of rank $P=N-R$. Even in this
case the expression (\ref{Mdt_B=PP_dt}) holds, and from this result it is
quite simple to verify that if $\lambda^{^{(\alpha)}}$ are $R$ eigenvectors of
$M^{AB}$, $M^{AB}\lambda_{A}^{^{(\alpha)}}=0$ then
\[
\frac{\partial V}{\partial z_{A}}\lambda_{A}^{^{(\alpha)}}dt_{0}=0.
\]
Since $M^{AB}$ has rank $P=N-R$ then there is a submatrix $P\mathsf{x}P$ of
$M^{AB}$ such that%
\[
\det\left(  M^{ab}\right)  \neq
0,\,\,\,\,\,\,\,\,\,\,\,\,\,\,\,\,\,\,\,a,b=1,...,P,
\]
which implies in the existence of $M_{ab}^{-1}$. Hence we can rewrite
(\ref{Mdt_B=PP_dt}) as (considering $a=1,...,P;\,\alpha=P+1,...,N$)
\[
-\left\{  \phi^{A},\phi^{0}\right\}  dt_{0}=M^{AB}dt_{B}=M^{Ab}dt_{b}%
+M^{A\beta}dt_{\beta},\Rightarrow
\]%
\begin{equation}
\Rightarrow-\left\{  \phi^{A},\phi^{0}\right\}  dt_{0}=\delta_{a}^{A}%
M^{ab}dt_{b}+\delta_{\alpha}^{A}M^{\alpha b}dt_{b}+\delta_{a}^{A}M^{a\beta
}dt_{\beta}+\delta_{\alpha}^{A}M^{\alpha\beta}dt_{\beta}. \label{PP_deltas}%
\end{equation}
Taking $A=a$ we see some $dt_{b}$ can be expressed as a linear combinations of
$dt_{\beta}$ and $dt$:%
\[
dt_{b}=-M_{ba}^{-1}\left[  \left\{  \phi^{a},\phi^{0}\right\}  dt_{0}%
+M^{a\beta}dt_{\beta}\right]  \Rightarrow
\]%
\begin{equation}
\Rightarrow dt_{b}=-M_{ba}^{-1}\left\{  \phi^{a},\phi^{\beta^{\prime}%
}\right\}  dt_{\beta^{\prime}},\,\,\,\,\,\,\,\,\,\,\,\beta^{\prime}=\left\{
0,\beta\right\}  . \label{d_t_bar}%
\end{equation}

If we consider now the case $A=\alpha$ it follows%
\[
M^{\alpha b}dt_{b}=-M^{\alpha\beta}dt_{\beta}-\left\{  \phi^{\alpha},\phi
^{0}\right\}  dt,
\]
and if we use (\ref{d_t_bar}) and consider $dt_{\beta}$and $dt_{0}$ as
independent parameters, then%
\begin{equation}
\left\{
\begin{array}
[c]{c}%
\left\{  \phi^{\alpha},\phi^{0}\right\}  =\left\{  \phi^{\alpha},\phi
^{b}\right\}  M_{ba}^{-1}\left\{  \phi^{a},\phi^{0}\right\} \\
\left\{  \phi^{\alpha},\phi^{\beta}\right\}  =\left\{  \phi^{\alpha},\phi
^{b}\right\}  M_{ba}^{-1}\left\{  \phi^{a},\phi^{\beta}\right\}
\end{array}
\right.  , \label{fix_sub_esp}%
\end{equation}
which tell us that, if (\ref{d_t_bar}) is satisfied, $\left\{  \phi^{\alpha
},\phi^{0}\right\}  $ and the elements $\left\{  \phi^{\alpha},\phi^{\beta
}\right\}  $ of the matrix $M^{AB}=\left\{  \phi^{A},\phi^{B}\right\}  $ must
be related to $\left\{  \phi^{a},\phi^{0}\right\}  ,\,\left\{  \phi^{a}%
,\phi^{\beta}\right\}  ,$\thinspace$\left\{  \phi^{a},\phi^{b}\right\}  $. The
second expression is in agreement with the fact that $M^{AB}$ is singular,
while the first one must be faced as conditions that actually fix the subspace
of the parameters where the system can be integrable. It becomes clear from
the results above that the case $A=a$ brings information about the dependence
among the $P$ parameters $dt_{b}$ and the $R$ parameters $dt_{\beta^{\prime}}%
$, while the case $A=\alpha$ stablishes $R$ conditions that determine the
subspace of integrability.

With (\ref{d_t_bar}) the differential of $E=E\left(  z,p\right)  $ becomes%
\begin{align}
dE  &  =\left[  \left\{  E,\phi^{\beta^{\prime}}\right\}  -\left\{  E,\phi
^{b}\right\}  M_{ba}^{-1}\left\{  \phi^{a},\phi^{\beta}\right\}  \right]
dt_{\beta^{\prime}}\Rightarrow\nonumber\\
&  \Rightarrow dE=\left\{  E,\phi^{\beta^{\prime}}\right\}  _{\ast}%
dt_{\beta^{\prime}}, \label{dE_sing}%
\end{align}
where the Brackets $\left\{  F,G\right\}  _{\ast}$ are introduced%
\begin{equation}
\left\{  F,G\right\}  _{\ast}\equiv\left\{  F,G\right\}  -\left\{  F,\phi
^{b}\right\}  M_{ba}^{-1}\left\{  \phi^{a},G\right\}  . \label{PD_HJ_sing}%
\end{equation}

And if we consider $F=F\left(  z_{A}\right)  $ and $G=G\left(  z_{B}\right)
,$%
\[
\left\{  F,G\right\}  _{\ast}=\frac{\partial F}{\partial z_{a}}M_{ab}%
^{-1}\frac{\partial G}{\partial z_{b}},
\]
and for $F=z_{A}$ and $G=z_{B}$%
\[
\left\{  z_{A},z_{B}\right\}  _{\ast}=\delta_{A}^{a}M_{ab}^{-1}\delta_{B}%
^{b}\Rightarrow\left\{
\begin{array}
[c]{c}%
\left\{  z_{a},z_{b}\right\}  _{\ast}=M_{ab}^{-1},\\
\left\{  z_{a},z_{\beta}\right\}  _{\ast}=0,\\
\left\{  z_{\alpha},z_{\beta}\right\}  _{\ast}=0.
\end{array}
\right.  .
\]
These two last results show the existence of a reduced symplectic structure in
phase space of a singular $M^{AB}$.

Taking $E=z_{C}$ in (\ref{dE_sing}) it follows%
\begin{align*}
dz_{C}  &  =\left\{  F,\phi^{\beta^{\prime}}\right\}  _{\ast}dt_{\beta
^{\prime}}=\left[  \left\{  z_{C},\phi^{\beta^{\prime}}\right\}  -\left\{
z_{C},\phi^{b}\right\}  M_{ba}^{-1}\left\{  \phi^{a},\phi^{\beta^{\prime}%
}\right\}  \right]  dt_{\beta^{\prime}}=\\
&  =\left[  \delta_{C}^{\beta^{\prime}}-\left\{  z_{C},\phi^{b}\right\}
M_{ba}^{-1}\left\{  \phi^{a},\phi^{\beta^{\prime}}\right\}  \right]
dt_{\beta^{\prime}}=\\
&  =\left[  \delta_{C}^{\beta^{\prime}}-\delta_{C}^{b}M_{ba}^{-1}\left\{
\phi^{a},\phi^{\beta^{\prime}}\right\}  \right]  dt_{\beta^{\prime}}%
\end{align*}
and for $C=\beta^{\prime}$:%
\[
dz_{\beta^{\prime}}=dt_{\beta^{\prime}},
\]
that shows $z_{\beta^{\prime}}$ remain arbitrary parameters in this
construction. For $C=b$,%
\[
dz_{b}=-M_{ba}^{-1}\left\{  \phi^{a},\phi^{\beta^{\prime}}\right\}
dt_{\beta^{\prime}},
\]
which is consistent with (\ref{d_t_bar}) since $z_{b}=t_{b}$. This expression
can still be written as
\[
dz_{b}=M_{ba}^{-1}\left[  \frac{\partial K^{a}}{\partial z_{\beta^{\prime}}%
}-\frac{\partial K^{\beta^{\prime}}}{\partial z_{a}}\right]  dt_{\beta
^{\prime}},
\]
when we explicit the Poisson Bracket $\left\{  \phi^{A},\phi^{\beta^{\prime}%
}\right\}  $.

\section{Examples}

In order to ilustrate how the method works, let us consider the two examples
studied in \cite{Guler -lin-veloc}, where the HJ method was also applied.

\textbf{1}- Starting with the Lagrangian
\begin{gather*}
L=\left(  z_{2}+z_{3}\right)  \dot{z}_{1}+z_{4}\dot{z}_{3}+W\left(
z_{2},z_{3},z_{4}\right)  ,\\
W\left(  z_{2},z_{3},z_{4}\right)  =\frac{1}{2}\left[  \left(  z_{4}\right)
^{2}-2z_{2}z_{3}-\left(  z_{3}\right)  ^{2}\right]  ,
\end{gather*}
in a four dimensional (coordinate) space, it is immediate to identify $K^{A}$
and the constraints $\phi^{A}$:%
\[
\left\{
\begin{array}
[c]{l}%
K^{1}=z_{2}+z_{3},\\
K^{2}=0,\\
K^{3}=z_{4},\\
K^{4}=0,\\
V=-W,
\end{array}
\right.  \Rightarrow\left\{
\begin{array}
[c]{l}%
\phi^{1}=p^{1}-K^{1}=p^{1}-\left(  z_{2}+z_{3}\right)  =0,\\
\phi^{2}=p^{2}-K^{2}=p^{2}=0,\\
\phi^{3}=p^{3}-K^{3}=p^{3}-z_{4}=0,\\
\phi^{4}=p^{4}-K^{4}=p^{4}=0,\\
\phi^{0}=p^{0}+V=p^{0}-\frac{1}{2}\left[  \left(  z_{4}\right)  ^{2}%
-2z_{2}z_{3}-\left(  z_{3}\right)  ^{2}\right]  =0.
\end{array}
\right.
\]
From here, the matrix $M^{AB}$ is given by%
\[
M^{AB}=\frac{\partial K^{B}}{\partial z_{A}}-\frac{\partial K^{A}}{\partial
z_{B}}=\left\{  \phi^{A},\phi^{B}\right\}  \Rightarrow\left(  M^{AB}\right)
=\left(
\begin{array}
[c]{cccc}%
0 & -1 & -1 & 0\\
1 & 0 & 0 & 0\\
1 & 0 & 0 & -1\\
0 & 0 & 1 & 0
\end{array}
\right)  ,
\]
which is regular and whose inverse is%
\[
\left(  M_{AB}^{-1}\right)  =\left(
\begin{array}
[c]{cccc}%
0 & 1 & 0 & 0\\
-1 & 0 & 0 & -1\\
0 & 0 & 0 & 1\\
0 & 1 & -1 & 0
\end{array}
\right)  .
\]
Now we can find the equations of motion by constructing the generalized
brackets and using (\ref{dE}), or using (\ref{eq_mov_reg}) in a
straightforward way:%
\[
\left\{
\begin{array}
[c]{l}%
dz_{1}=z_{3}dt,\\
dz_{2}=z_{4}dt,\\
dz_{3}=-z_{4}dt,\\
dz_{4}=-z_{2}dt,
\end{array}
\right.  \Rightarrow\left\{
\begin{array}
[c]{l}%
\dot{z}_{1}=z_{3},\\
\dot{z}_{2}=z_{4},\\
\dot{z}_{3}=-z_{4},\\
\dot{z}_{4}=-z_{2}.
\end{array}
\right.
\]
Manipulating the second and fourth equations we see that%
\[
\ddot{z}_{2}+z_{2}=0\Rightarrow z_{2}=-A\cos t+B\sin t,
\]
and by direct substitution into the second equation it follows
\[
z_{4}=A\sin t+B\cos t.
\]
By integration it is verified that%
\begin{align*}
z_{3}  &  =A\cos t-B\sin t+C,\\
z_{1}  &  =A\sin t+B\cos t+Ct+D.
\end{align*}
Now, substituting these results in the constraints we can obtain the momenta:%
\[
\left\{
\begin{array}
[c]{l}%
p^{1}=C,\\
p^{2}=0,\\
p^{3}=A\sin t+B\cos t,\\
p^{4}=0,
\end{array}
\right.
\]
and the problem is completely solved since we know all phase space
variables.\ Comparing these results with those obtained in \cite{Guler
-lin-veloc} we see some differences, the main one being the linear dependence
of $z_{1}$ with $t$. In fact we verify that the result of \cite{Guler
-lin-veloc} is a particular case of the one obtained here.

\textbf{2}- Let us now consider the Lagrangian
\begin{gather*}
L=\left(  z_{2}+z_{3}\right)  \dot{z}_{1}+k\dot{z}_{3}+W\left(  z_{2}%
,z_{3}\right)  ,\\
W\left(  z_{2},z_{3},z_{4}\right)  =\frac{1}{2}\left[  k^{2}-2z_{2}%
z_{3}-\left(  z_{3}\right)  ^{2}\right]  ,
\end{gather*}
in a three dimensional (coordinate) space. We identify%
\[
\left\{
\begin{array}
[c]{l}%
K^{1}=z_{2}+z_{3},\\
K^{2}=0,\\
K^{3}=k,\\
V=-W,
\end{array}
\right.  \Rightarrow\left\{
\begin{array}
[c]{l}%
\phi^{1}=p^{1}-\left(  z_{2}+z_{3}\right)  =0,\\
\phi^{2}=p^{2}=0,\\
\phi^{3}=p^{3}-k=0,\\
\phi^{0}=p^{0}-\frac{1}{2}\left[  k^{2}-2z_{2}z_{3}-\left(  z_{3}\right)
^{2}\right]  =0,
\end{array}
\right.
\]
and construct $M^{AB}$:%
\[
\left(  M^{AB}\right)  =\left(
\begin{array}
[c]{ccc}%
0 & -1 & -1\\
1 & 0 & 0\\
1 & 0 & 0
\end{array}
\right)  .
\]
This is a singular matrix and has rank 2, and then we must find an inversible
submatrix $M^{ab}$; this can be done by choosing%
\[
\left(  M^{ab}\right)  =\left(
\begin{array}
[c]{cc}%
M^{11} & M^{13}\\
M^{31} & M^{33}%
\end{array}
\right)  =\left(
\begin{array}
[c]{cc}%
0 & -1\\
1 & 0
\end{array}
\right)  \Rightarrow\left(  M_{ab}^{-1}\right)  =\left(
\begin{array}
[c]{cc}%
0 & 1\\
-1 & 0
\end{array}
\right)  .
\]
This choice implies that $t_{\beta}=z_{2}$ ($\phi^{\alpha}=\phi^{2}$) and
$t_{b}=\left\{  z_{1},z_{3}\right\}  $ ($\phi^{a}=\left\{  \phi^{1},\phi
^{3}\right\}  $), and the construction of the generalized brackets leads us to
the following equations of motion:%
\[
\left\{
\begin{array}
[c]{l}%
dz_{1}=\left(  z_{2}+z_{3}\right)  dt,\\
dz_{3}=-dz_{2}.
\end{array}
\right.
\]
By direct integration of the second equation it follows%
\[
z_{3}=-z_{2}+C;
\]
which shows us that%
\[
z_{1}=Ct+D.
\]
Now we must look for the condition that fixes the subspace where the system is
integrable:%
\begin{align*}
\left\{  \phi^{2},\phi^{0}\right\}   &  =\left\{  \phi^{2},\phi^{b}\right\}
M_{ba}^{-1}\left\{  \phi^{a},\phi^{0}\right\}  \Rightarrow\\
&  \Rightarrow-z_{3}=\left(
\begin{array}
[c]{cc}%
1 & 0
\end{array}
\right)  \left(
\begin{array}
[c]{cc}%
0 & 1\\
-1 & 0
\end{array}
\right)  \left(
\begin{array}
[c]{c}%
0\\
-\left(  z_{2}+z_{3}\right)
\end{array}
\right)  \Rightarrow z_{2}=0.
\end{align*}
We see that this system is integrable in the subspace where $z_{2}=0$, which
leads to conclude that $z_{3}=C$. Substituting these results in the
constraints, it follows%
\[
\left\{
\begin{array}
[c]{l}%
p^{1}=C,\\
p^{2}=0,\\
p^{3}=k.
\end{array}
\right.
\]
The problem is then completely solved.\ If we now compare these results with
those obtained in \cite{Guler -lin-veloc} we see that they are in agreement if
we correctly fix the values of $C_{1}$,\ $C_{2}$, and $C_{3}$\ of reference
\cite{Guler -lin-veloc}.

\textbf{3} - Now we will consider a third example of a system of fields known
as Proca's model. The Lagrangian density considered here has the Palatini's
form, where the fields $A_{\mu}\left(  x\right)  $ and $F^{\mu\nu}\left(
x\right)  (\mu,\nu=0,1,2,3;\,i,j=1,2,3)$\ are considered as independent fields
($z_{A}=A_{\mu},F^{\mu\nu}$):%
\begin{align*}
\mathcal{L}  &  =\frac{1}{4}A_{\nu}\partial_{\mu}\left(  F^{\mu\nu}-F^{\nu\mu
}\right)  -\frac{1}{4}F^{\mu\nu}\left(  \partial_{\mu}A_{\nu}-\partial_{\nu
}A_{\mu}\right)  +\frac{1}{4}F^{\mu\nu}F_{\mu\nu}+\frac{1}{2}m^{2}A_{\mu
}A^{\mu}\\
&  =-\frac{1}{4}\left(  F^{0\nu}-F^{\nu0}\right)  \partial_{0}A_{\nu}+\frac
{1}{4}A_{i}\partial_{0}F^{0i}-\frac{1}{4}A_{i}\partial_{0}F^{i0}-\mathcal{H},
\end{align*}
where%
\[
\mathcal{H}=-\frac{1}{4}F^{\mu\nu}F_{\mu\nu}+\frac{1}{4}\left(  F^{i\nu
}-F^{\nu i}\right)  \partial_{i}A_{\nu}-\frac{1}{4}A_{\nu}\partial_{i}\left(
F^{i\nu}-F^{\nu i}\right)  -\frac{1}{2}m^{2}A_{\mu}A^{\mu}.
\]
We then identify%
\begin{gather*}
\left\{
\begin{array}
[c]{l}%
K^{\nu}\left(  x\right)  =-\frac{1}{4}\left(  F^{0\nu}\left(  x\right)
-F^{\nu0}\left(  x\right)  \right)  ,\\
K_{00}\left(  x\right)  =0,\\
K_{0i}\left(  x\right)  =\frac{1}{4}A_{i}\left(  x\right)  ,\\
K_{i0}\left(  x\right)  =-\frac{1}{4}A_{i}\left(  x\right)  ,\\
K_{ij}\left(  x\right)  =0\\
V\left(  x\right)  =\mathcal{H}\left(  x\right)  ,
\end{array}
\right.  \Rightarrow\\
\Rightarrow\left\{
\begin{array}
[c]{l}%
\phi^{\nu}\left(  x\right)  =\pi^{\nu}\left(  x\right)  +\frac{1}{4}\left(
F^{0\nu}\left(  x\right)  -F^{\nu0}\left(  x\right)  \right)  =0,\\
\phi_{00}\left(  x\right)  =\Pi_{00}\left(  x\right)  =0,\\
\phi_{0i}\left(  x\right)  =\Pi_{0i}\left(  x\right)  -\frac{1}{4}A_{i}\left(
x\right)  =0,\\
\phi_{i0}\left(  x\right)  =\Pi_{i0}\left(  x\right)  +\frac{1}{4}A_{i}\left(
x\right)  =0,\\
\phi_{ij}\left(  x\right)  =\Pi_{ij}\left(  x\right)  =0,\\
\phi^{t}\left(  x\right)  =p^{0}+\mathcal{H}\left(  x\right)  =0.
\end{array}
\right.
\end{gather*}
The $M^{AB}$ matrix, now defined as
\[
M^{A_{x}B_{y}}=\frac{\delta K^{B}\left(  y\right)  }{\delta z_{A}\left(
x\right)  }-\frac{\delta K^{A}\left(  x\right)  }{\delta z_{B}\left(
y\right)  },
\]
is%
\[
\left(  M^{A_{x}B_{y}}\right)  =\left(
\begin{array}
[c]{ccccc}%
0 & 0 & \frac{1}{2}\delta_{j}^{\mu}\delta\left(  x-y\right)  & -\frac{1}%
{2}\delta_{j}^{\mu}\delta\left(  x-y\right)  & 0\\
0 & 0 & 0 & 0 & 0\\
-\frac{1}{2}\delta_{i}^{\nu}\delta\left(  x-y\right)  & 0 & 0 & 0 & 0\\
\frac{1}{2}\delta_{i}^{\nu}\delta\left(  x-y\right)  & 0 & 0 & 0 & 0\\
0 & 0 & 0 & 0 & 0
\end{array}
\right)  ,
\]
and it is not inversible. So we must find an inversible submatrix $M^{ab}$,
what can be done by choosing $t_{\beta}\left(  x\right)  =\left\{
A_{0}\left(  x\right)  ,F^{00}\left(  x\right)  ,F^{j0}\left(  x\right)
,F^{ij}\left(  x\right)  \right\}  $ and $t_{b}=\left\{  A_{i}\left(
x\right)  ,F^{0j}\left(  x\right)  \right\}  $, such that%
\begin{align*}
\left(  M^{a_{x}b_{y}}\right)   &  =\left(
\begin{array}
[c]{cc}%
\mathbf{0}_{3\times3} & \mathbf{1}_{3\times3}\\
-\mathbf{1}_{3\times3} & \mathbf{0}_{3\times3}%
\end{array}
\right)  \frac{1}{2}\delta\left(  x-y\right)  \Rightarrow\\
&  \Rightarrow\left(  M_{c_{z}a_{x}}^{-1}\right)  =\left(
\begin{array}
[c]{cc}%
\mathbf{0}_{3\times3} & -\mathbf{1}_{3\times3}\\
\mathbf{1}_{3\times3} & \mathbf{0}_{3\times3}%
\end{array}
\right)  2\delta\left(  z-x\right)
\end{align*}

In order to construct the generalized brackets we must calculate $\left\{
\phi^{a_{y}},\phi^{\beta_{x}^{\prime}}\right\}  $:%
\begin{align*}
&
\begin{array}
[c]{c}%
\left\{  \phi^{i}\left(  y\right)  ,\phi^{0}\left(  x\right)  \right\}
=M^{i_{y}\left(  00\right)  _{x}}=0,\\
\left\{  \phi^{i}\left(  y\right)  ,\phi_{00}\left(  x\right)  \right\}
=M^{i_{y}\left(  00\right)  _{x}}=0,\\
\left\{  \phi^{i}\left(  y\right)  ,\phi_{j0}\left(  x\right)  \right\}
=M^{i_{y}\left(  j0\right)  _{x}}=-\frac{1}{2}\delta_{j}^{i}\delta\left(
x-y\right)  ,\\
\left\{  \phi^{i}\left(  y\right)  ,\phi_{mn}\left(  x\right)  \right\}
=M^{i_{y}\left(  mn\right)  _{x}}=0,
\end{array}
\\
&
\begin{array}
[c]{c}%
\left\{  \phi_{0i}\left(  y\right)  ,\phi^{0}\left(  x\right)  \right\}
=M^{\left(  0i\right)  _{y}0_{y}}=0,\\
\left\{  \phi_{0i}\left(  y\right)  ,\phi_{00}\left(  x\right)  \right\}
=M^{\left(  0i\right)  _{y}\left(  00\right)  _{x}}=0,\\
\left\{  \phi_{0i}\left(  y\right)  ,\phi_{j0}\left(  x\right)  \right\}
=M^{\left(  0i\right)  _{y}\left(  j0\right)  _{x}}=0,\\
\left\{  \phi_{0i}\left(  y\right)  ,\phi_{mn}\left(  x\right)  \right\}
=M^{\left(  0i\right)  _{y}\left(  mn\right)  _{x}}=0,
\end{array}
\end{align*}%
\begin{align*}
\left\{  \phi^{i}\left(  y\right)  ,\phi^{t}\left(  x\right)  \right\}   &
=-\frac{\partial\mathcal{H}\left(  x\right)  }{\partial A_{i}\left(  y\right)
}=-\frac{1}{4}\left[  F^{ji}\left(  x\right)  -F^{ij}\left(  x\right)
\right]  \partial_{j}^{x}\delta\left(  x-y\right)  +\\
&  +\frac{1}{4}\left[  \partial_{j}F^{ji}\left(  x\right)  -\partial_{j}%
F^{ij}\left(  x\right)  \right]  \delta\left(  x-y\right)  +m^{2}A^{i}\left(
x\right)  \delta\left(  x-y\right)  ,\\
\left\{  \phi_{0i}\left(  y\right)  ,\phi^{t}\left(  x\right)  \right\}   &
=-\frac{\partial\mathcal{H}\left(  x\right)  }{\partial F^{0i}\left(
y\right)  }=\left[  \frac{1}{2}F_{0i}\left(  x\right)  +\frac{1}{4}%
\partial_{i}A_{0}\left(  x\right)  \right]  -\frac{1}{4}A_{0}\left(  x\right)
\partial_{i}^{x}\delta\left(  x-y\right)  .
\end{align*}
From this results we see that for any function $G=G\left(  z,p\right)  $\
\begin{align}
dG &  =\left\{  G,\phi^{\beta_{x}^{\prime}}\right\}  dt_{\beta_{x}^{\prime}%
}+\nonumber\\
&  -\left\{  G,\phi^{c_{z}}\right\}  \left(  M_{c_{z}a_{y}}^{-1}\right)
\left\{  \phi^{a_{y}},\phi^{t}\left(  x\right)  \right\}  dt\left(  x\right)
+\nonumber\\
&  -\left\{  G,\phi^{c_{z}}\right\}  \left(  M_{c_{z}a_{y}}^{-1}\right)
\left\{  \phi^{a_{y}},\phi_{j0}\left(  x\right)  \right\}  dF^{j0}\left(
x\right)  .\label{dG_Proca}%
\end{align}

Before obtaining the field equations we will look for the conditions that fix
the subspace where the system is integrable,
\[
\left\{  \phi^{\alpha_{w}},\phi^{t}\left(  x\right)  \right\}  =\left\{
\phi^{\alpha_{w}},\phi^{b_{z}}\right\}  M_{b_{z}a_{y}}^{-1}\left\{
\phi^{a_{y}},\phi^{t}\left(  x\right)  \right\}  .
\]
For $\phi^{\alpha_{w}}=\phi^{0}\left(  w\right)  $,
\begin{align*}
\left\{  \phi^{0}\left(  w\right)  ,\phi^{t}\left(  x\right)  \right\}   &
=-\frac{1}{4}\left[  F^{j0}\left(  x\right)  -F^{0j}\left(  x\right)  \right]
\partial_{j}^{x}\delta\left(  x-w\right)  +\\
&  +\frac{1}{4}\left[  \partial_{j}F^{j0}\left(  x\right)  -\partial_{j}%
F^{0j}\left(  x\right)  \right]  \delta\left(  x-w\right)  +\\
&  +m^{2}A^{0}\left(  x\right)  \delta\left(  x-w\right)  ,
\end{align*}
and%
\[
\left\{  \phi^{0}\left(  w\right)  ,\phi^{b_{z}}\right\}  M_{b_{z}a_{y}}%
^{-1}\left\{  \phi^{a_{y}},\phi^{t}\left(  x\right)  \right\}  =0,
\]
which lead us to conclude that%
\begin{equation}
\frac{1}{2}\left[  \partial_{j}F^{j0}\left(  x\right)  -\partial_{j}%
F^{0j}\left(  x\right)  \right]  +m^{2}A^{0}\left(  x\right)  =0.\label{cond1}%
\end{equation}
Considering now $\phi^{\alpha_{w}}=\phi_{00}\left(  w\right)  $,%
\[
\left\{  \phi_{00}\left(  w\right)  ,\phi^{t}\left(  x\right)  \right\}
=\frac{1}{2}F_{00}\left(  x\right)  \delta\left(  w-x\right)  ,
\]%
\[
\left\{  \phi_{00}\left(  w\right)  ,\phi^{\bar{B}_{z}}\right\}  M_{\bar
{B}_{z}\bar{A}_{y}}^{-1}\left\{  \phi^{\bar{A}_{y}},\phi^{t}\left(  x\right)
\right\}  =0,
\]%
\begin{equation}
\therefore F_{00}\left(  x\right)  =0.\label{cond2}%
\end{equation}
For $\phi^{\alpha_{w}}=\phi_{ij}\left(  w\right)  $,%
\[
\left\{  \phi_{ij}\left(  w\right)  ,\phi^{t}\left(  x\right)  \right\}
=\frac{1}{2}\left[  F_{ij}\left(  x\right)  -\left(  \partial_{i}A_{j}\left(
x\right)  -\partial_{j}A_{i}\left(  x\right)  \right)  \right]  \delta\left(
w-x\right)  ,
\]%
\[
\left\{  \phi_{ij}\left(  w\right)  ,\phi^{\bar{B}_{z}}\right\}  M_{\bar
{B}_{z}\bar{A}_{y}}^{-1}\left\{  \phi^{\bar{A}_{y}},\phi^{t}\left(  x\right)
\right\}  =0,
\]%
\begin{equation}
\therefore F_{ij}\left(  x\right)  =\partial_{i}A_{j}\left(  x\right)
-\partial_{j}A_{i}\left(  x\right)  .\label{cond3}%
\end{equation}
At last, considering $\phi^{\alpha_{w}}=\phi_{j0}\left(  w\right)  $,%
\[
\left\{  \phi_{j0}\left(  w\right)  ,\phi^{t}\left(  x\right)  \right\}
=\frac{1}{2}\left[  F_{j0}\left(  x\right)  -\partial_{j}A_{0}\left(
x\right)  \right]  \delta\left(  w-x\right)  ,
\]%
\[
\left\{  \phi_{j0}\left(  w\right)  ,\phi^{\bar{B}_{z}}\right\}  M_{\bar
{B}_{z}\bar{A}_{y}}^{-1}\left\{  \phi^{\bar{A}_{y}},\phi^{t}\left(  x\right)
\right\}  =-\frac{1}{2}\left[  F_{0j}\left(  x\right)  -\partial_{j}%
A_{0}\left(  x\right)  \right]  \delta\left(  w-x\right)  ,
\]%
\begin{equation}
\therefore F_{j0}\left(  x\right)  =-F_{0j}\left(  x\right)  .\label{cond4}%
\end{equation}

Now we are able to obtain the field equations. Taking $G=A_{j}\left(
w\right)  $ in (\ref{dG_Proca}) we have%
\begin{equation}
dA_{j}\left(  w\right)  =\left[  F_{0j}\left(  w\right)  +\partial_{j}%
A_{0}\left(  w\right)  \right]  dt. \label{eq_mov_Proca1}%
\end{equation}

If we now consider $G=F^{0j}\left(  w\right)  $,%
\begin{equation}
d\left[  F^{0j}\left(  w\right)  -F^{j0}\left(  w\right)  \right]  =\left[
\partial_{m}F^{jm}\left(  w\right)  -\partial_{m}F^{mj}\left(  w\right)
-2m^{2}A^{k}\left(  w\right)  \right]  dt. \label{eq_mov_Proca2}%
\end{equation}

The results above (eqs. (\ref{cond1}-\ref{eq_mov_Proca2}))can be summarized as%
\begin{gather*}
F_{\mu\nu}\left(  x\right)  =-F_{\nu\mu}\left(  x\right)  =\partial_{\mu
}A_{\nu}\left(  x\right)  -\partial_{\nu}A_{\mu}\left(  x\right)  ,\\
\partial_{\mu}F^{\mu\nu}\left(  x\right)  +m^{2}A^{\nu}\left(  x\right)  =0.
\end{gather*}
Moreover, when we take the divergence of this last result, as consequence of
the antisymmetry property of $F_{\mu\nu}$, it follows:%
\[
\partial_{\nu}A^{\nu}\left(  x\right)  =0.
\]
When we compare these results with those obtained in reference \cite{coreanos}%
, where an analysis of the Proca model is conducted with the usual
Hamilton-Jacobi formalism, the agreement is manifest.

\section{Final Remarks}

In this work we have studied how systems described by Lagrangians with linear
velocities can be treated in Hamilton-Jacobi formalism. Initially we observed
that all \emph{momenta} were constrained and therefore all coordinates had the
status of parameters. Then, after applying integrability conditions, we saw
that if the conditions $\left\{  \phi^{A},\phi^{0}\right\}  =0,$ $M^{AB}=0$
are satisfied, then the system of total differential equation is integrable,
and all the parameters are independent.

But, if these conditions are not satisfied then the system of differential
equation can be integrable only in a subspace of the phase space. Two distinct
cases were analysed. Firstly we considered the $M^{AB}$regular case, where the
system is integrable in the subspace where all the parameters are
time-dependent (see (\ref{dt_B})). In this subspace we verified that new
generalized brackets could be introduced, which allowed us to recognize the
symplectic structure of phase space in HJ approach.

Secondly we considered the $M^{AB}$ singular case, and we saw that
integrability is achieved in a subspace where $t_{0}$ and $t_{\beta}$ are
independent parameters. We also verified that some extra conditions must be
satisfied in order that this subspace could be determined. In this subspace
generalized brackets were also introduced and the symplectic structure could
also be recognized. Here one interesting feature can be pointed out. In this
case we can separate the parameters $t_{B}$ in two distinct sets: one is
composed by the parameters $t_{\beta}$, and the other by $t_{b}$, which are
the real dynamical variables of the problem (in example 1, all $z_{A},$
$A=1,...,4,$ are dynamical variables, while in example 2 and 3, only $z_{1}$,
$z_{3}$ and $A_{i}\left(  x\right)  $, $F^{0j}\left(  x\right)  $ are
dynamical, respectively). In the same way, the associated constraints can also
be separated in two sets composed by $\phi^{b}$ and $\phi^{\beta}$,
respectively. The interesting feature of this separation is to identify the
constraints that allow one to construct the inversible submatrix $M^{ab}$,
used to define the generalized brackets.

Moreover we believe the introduction of generalized brackets can be done not
only in systems described by first order actions, but also in any system which
has a non-null matrix $M^{AB}=\left\{  \phi^{A},\phi^{B}\right\}  $ with an
inversible submatrix $M^{ab}$. This case is still under consideration by the authors.

At last we must notice that, although we considered only usual variables in
this work, the extension to treat berezinian variables is quite immediate.

\bigskip

\textbf{Acknowledgements}

\bigskip

The authors would like to thanks the referee for his comments, which allowed
them to improve their work. M. C. Bertin thanks CNPq for full support; B.
M.\ Pimentel thanks CNPq and FAPESP,\ (grant number 02/00222-9) for partial
support; P. J. Pompeia thanks the staff of CTA for incentive and support.

\end{document}